\documentclass[11 pt, a4paper]{article}
\usepackage{epsfig}
\usepackage{epsf}
\usepackage{bm}                 % for bold math symbols
\usepackage{amsmath}
\usepackage{amssymb}
\usepackage{dcolumn}
\usepackage{graphicx}
\usepackage{psfrag}
\usepackage{latexsym}
\usepackage{color}

\newcommand{\beq}{\begin{equation}}
\newcommand{\eeq}{\end{equation}}
\newcommand{\beqa}{\begin{eqnarray}}
\newcommand{\eeqa}{\end{eqnarray}}

\def\Symp#1,#2,#3,#4.{\left[\left(\begin{array}{c}#1\\#2\end{array}\right),\left(\begin{array}{c}#3\\#4\end{array}\right)\right]}
\def\Vec#1,#2.{\left(\!\begin{array}{c}#1\\#2\end{array}\!\right)}
\def\vec#1,#2.{{#1\choose{#2}}}

\newcommand{\bx}{{\bf x}}
\newcommand{\by}{{\bf y}}

\definecolor{redcom}{rgb}{1,0.1,0.2}
\definecolor{querycol}{rgb}{0.2,0.2,1}

\definecolor{purplerep}{rgb}{1,0.1,1}

\definecolor{green}{rgb}{0.1,0.8,1}

\begin{document}

\title{Generalized guidance equation for peaked quantum solitons: the single
 particle case\footnote{To appear in the annales de la Fondation de Broglie in 2017, under the title ``de Broglie double solution and self-gravitation.''}}
\date{}
\author{}
\maketitle
\centerline{Thomas Durt\footnote{ Aix  Marseille  Universit\'e,  CNRS,
Centrale  Marseille, Institut Fresnel UMR 7249,13013 Marseille, France.email: thomas.durt@centrale-marseille.fr}}

% Force line breaks with \\

\abstract{  We study the Schr\"odinger-Newton equation (a generalisation of the linear Schr\"odinger equation, which contains a self-focusing  non-linear interaction of gravitational nature) in the light of de Broglie's double solution program. In particular we consider solutions of the Schr\"odinger-Newton equation which obey the so-called factorisation ansatz  \cite{new} according to which the full wave function is the product of a smoothly varying function with a peaked self-collapsed soliton. We show that these solitons obey a generalized de Broglie-Bohm guidance equation where the smooth function plays the role of the pilot-wave. We derive an Ehrenfest-like theorem for the Schr\"odinger-Newton equation and conjecture the existence of a stochastic subquantum medium in order to explain departures from classical trajectories.}

\section{Introduction}
{\bf Self-gravitational interaction.}

The so-called Schr\"odinger-Newton (S-N) equation\footnote{This equation is also often referred to as the (attractive) Schr\"odinger-Poisson equation \cite{Brezzi, Illner, Arriola} or the gravitational Schr\"odinger equation \cite{Jones}. } \cite{Jones} reads

\begin{equation}
{i}\hbar\frac{\partial\Psi(t,{{\bf x}})}{\partial t}=-\hbar^2\frac{\Delta\Psi(t,{{\bf x}})}{2m}
-Gm^2\int {d}^3 x'(\frac{|\Psi(t,{{\bf x}'})|^2}{|{{\bf x} -x'}|})\Psi(t,{{\bf x}}),\label{NS}
\end{equation}where $G$ represents the Newton gravitational constant and $m$ the mass of a quantum object.
It has been intensively studied in the past, due to the self-focusing character of the non-linear potential which could possibly explain the wave function collapse, seen in this context as a self-localisation process \cite{diosi84,penrose}. 

Our prior motivation is to study the manifestations of self-gravitation in presence of an external potential $V^L$ where $V^L$ represents the external potentials that are commonly considered when solving the linear Schr\"odinger equation (for instance electro-magnetic potentials). Contrary to the self-gravitational interaction which non-linearily depends on $\Psi$, $V^L$ does not depend on $\Psi$. It is thus represented by a self-adjoint operator, linearily acting on the Hilbert space, as usually.

We shall thus assume that at the same time an external potential $V^{L}$ acts on the particle together with a non-linear self-focusing potential $V^{NL}$ of gravitational nature: 
\begin{equation}
{i}\hbar\frac{\partial\Psi(t,{{\bf x}})}{\partial t}=-\hbar^2\frac{\Delta\Psi(t,{{\bf x}})}{2m}
+V^{L}(t,{{\bf x}})\Psi(t,{{\bf x}})+V^{NL}(\Psi)\Psi(t,{{\bf x}}),\label{nonfreeNL}
\end{equation}
with $V^{NL}(\Psi)=-Gm^2\int {d}^3 x'(\frac{|\Psi(t,{{\bf x}'})|^2}{|{{\bf x} -{{\bf x'}}}|})$, in the case of elementary particles\footnote{\label{coco}It is worth noting that other choices of non-linear self-interaction are possible \cite{Fargue,old,CDW}, which lead to essentially the same results as those derived in the present paper. For instance, in the case of an homogeneous sphere, whenever the mean width of the center-of-mass wave function is small enough in comparison to the size of the sphere, the gravitational self-interaction reduces, in a first approximation, to a non-linear harmonic potential (see \cite{diosi84,Chen,CDW}). Indeed,   $-Gm^2/|{{\bf x} -{{\bf x'}}}|$ must be replaced in this case by

$-G({M\over {4\pi R^3\over 3}})^2\int_{|\tilde x| \leq R, |\tilde x'| \leq R} {d}^3 \tilde x {d}^3 	\tilde x'\frac{1}{|{{\bf x}}_{CM} +\tilde{{\bf x}}-({{\bf x}}_{CM}'+\tilde {{\bf x}}')|}$

$\approx {GM^2 \over R}(-\frac{6}{5}+\frac{1}{2}(\frac{|{{\bf x}}_{CM}-{{\bf x}}'_{CM}|}{R})^2+{\cal O}((\frac{|{{\bf x}}_{CM}-{{\bf x}}'_{CM}|}{R})^3))$.}.

%Looking for static solutions of the many particles NS equation, one obtains the following equation for the ground state wave function of the center of mass, where $\bx$=$\bx_{CM}$ :
%\begin{equation}\frac{\hbar^2}{2M} \Delta\varphi(\bx) - {G M^2\over 2R^3}\! \int\!d^3y ~\!\big|\varphi(\by)\big|^2~|\bx-\by|^2~\!\varphi(\bx) = -\cE^D\varphi(\bx)~.\label{eqDiosieff}\end{equation}
%Here we introduced the effective parameter  $\cE^D$ as
%\begin{equation}\cE^D= {6GM^2\over 5R} - \cE~\!,\label{EpsD}\end{equation}
%with respect to the parameter $\cE$ used in the reduction \eqref{redtoChoq}. This treatment is only valid, of course, to the extent that Di{\'o}si's approximation to a harmonic potential is valid, i.e. for widths of the ground state that are quite smaller than the radius $R$ of the sphere.

If, for instance, we consider an electron, the self-gravitational potential is usually considered to be very weak and treated as a perturbation. The self-collapsed ground state (\ref{Choquard}) of (\ref{NS}) is usually predicted (see details in appendix) to be normalized to unity, in which case its size of the order of $\hbar^2/Gm^3$ (more or less 10$^{32}$ meter in the case of an electron). The corresponding ground state energy (\ref{Choquard-energy}) is of the order of $G^2m^5/\hbar^2$ which is very small (for instance, in the case of an electron it is very small compared to the usual energies of electronic orbitals in an atom). Now, this result is obtained by assuming that the norm of the ground state is equal to unity, which, following de Broglie, we consider to be a superfetatory condition (as discussed with more detail in the last section). Making use of the well-known scaling properties of equation (\ref{NS}) (see the review paper by S. Colin, T.D. and R. Willox (same issue) and also \cite{CDW,vanMeter}) , it is actually possible to reduce arbitrarily the size of the self-collapsed ground state; at the same time, its norm will increase to plus infinity and its energy will tend to minus infinity.

{\bf Non-standard normalisation.}

The first non-standard ingredient of our work is that we choose not to normalize to unity the L$_2$ norm of the wave function. Instead, we impose to begin with that the size of the self-collapsed ground state is very small (of the order of the Schwarzschild radius $Gm/c^2$, thus of the order of 10$^{-57}$ meter in the case of an electron\footnote{We were led to this choice by studying certain implications of our work that go beyond the scope of the present paper \cite{new}. Making use of  the scaling properties of (\ref{NS}) \cite{CDW}, the norm of the ground state is of the order of $(mc^2\hbar^2/G^2m^5)^{1/3}\approx 10^{30}$ in the case of an electron.}). The ground state energy is then of the order of $-mc^2$. Contrary to the usual approach, in which external potentials are strong compared to the self-gravitational potential, in our case, the external potentials are supposedly weak, while the self-gravitational potential is strong and we treat the former as a perturbation. In particular the ground state will supposedly be very stable, because the stability analysis of the S-N equation in absence of external potential (\ref{NS}) shows that the non-linearity will inhibit the spreading unless the kinetic energy is at least of the order of the ground state energy \cite{Arriola,vanMeter,CDW}. We shall take for granted, without demonstration, that this is still approximately true in presence of an external potential $V^L$. Therefore we expect that in first approximation the state of the particle is, up to galilean boosts and translations, the static ground state, solution of (\ref{Choquard}). Moreover, the spectrum of negative energy solutions of (\ref{Choquard}) is discrete (see \cite{Bernstein} and appendix), so that at usual temperatures, the transitions to excited static self-collapsed states are frozen.

What we are looking for is thus a solution of (\ref{nonfreeNL}) in the form of a soliton (solitary wave) which, in first approximation, looks like the self-collapsed ground state of (\ref{NS}). In the rest of the paper we shall identify this highly concentrated peak of energy with the quantum particle itself, which is consistent with the aforementioned stability criterion: stability is menaced whenever an energy of the order of $mc^2$ is communicated to the system. It is worth noting that in our approach, contrary to the mainstream approach to self-localisation \cite{diosi84,penrose,CDW,arxiv}, the system is supposedly collapsed to begin with, since arbitrary long times.
This picture is reminiscent of the so-called de Broglie-Bohm (dB-B) causal interpretation of quantum mechanics. In particular, the fact that the solution is expected to have a very large norm and amplitude is reminiscent of de Broglie's double solution program\footnote{Louis de Broglie proposed in 1927 a realistic interpretation of the quantum theory in which particles are guided by the solution of the linear Schr\"odinger equation ($\Psi_L$), in accordance with the so-called guidance equation  \cite{debrogliebook,debroglieend}. The theory was generalised by David Bohm in 1952 \cite{bohm521,bohm522}. Certain ingredients of de Broglie's original idea disappeared in Bohm's formulation, in particular the double solution program, according to which the particle is associated to a wave $u$ distinct from the pilot-wave $\Psi_L$. This program was never fully achieved, $u$ being sometimes treated as a moving singularity \cite{vigier}, and sometimes as a solution $\phi_{NL}$ of a non-linear equation (see \cite{debroglieend,Fargue} and the papers Fargue, and of Colin, Durt and Willox, same issue).}, who wrote \cite{debrogliebook}

{\it`` ... a set of two coupled solutions of the wave equation: one, the $\Psi$ wave, definite in phase, but, because of the continuous character of its amplitude, having only a statistical and  subjective meaning; the other, the $u$ wave of the same phase as the $\Psi$ wave but with an amplitude having very large values around a point in space and which ($\cdots$) can be used to describe the particle objectively."} 

The fact that in our approach the particle, represented by $\phi_{NL}$,  has a very small size is reminiscent of Bohm's description of particles as material points.

Our approach is also reminiscent of Poincar\'e's attempts \cite{Poincar} to explain the stability of the electron in terms of an internal self-attraction (the so-called Poincar\'e pressure), aimed at counterbalancing Coulomb self-repulsion. In our model, self-gravitation plays the role of the Poincar\'e pressure, and it counterbalances the spread of the soliton that we identify with the quantum particle\footnote{Several physicists of the de Broglie school, Fer, Lochak, Andrade e Silva, Lochak and others developed in the past models mixing Poincar\'e and de Broglie views on the stability of particles (see for instance \cite{debroglieend,Fargue,IHP} and references therein as well as the papers of Fargue, Drezet and Colin, Durt and Willox, same issue)). These ideas are in a sense unavoidable whenever we try and describing particles as localised waves.}.

{\bf Factorization ansatz.}

In a first step, we tried to find a double solution {\it \`a la} de Broglie in the form of the sum of a wave function $\Psi_L$  (where $\Psi_L$ is a solution of the linear Schr\"odinger equation (\ref{S1V})) and of a soliton $\phi_{NL}$. However, due to the intrinsic non-linearity of (\ref{nonfreeNL}), we did not manage to derive interesting results.

This brings us to the second non-standard ingredient of our paper which is that we tried to solve (\ref{nonfreeNL}) with an ansatz solution $\Psi$ which factorizes (\ref{ansatz}) into the product of two functions  $\Psi_L$ and $\phi_{NL}$: 

\begin{equation}\Psi(t,{{\bf x}})=\Psi_L(t,{{\bf x}})\cdot \phi_{NL}(t,{{\bf x}}), \label{ansatz}\end{equation}

 for which we imposed that $\Psi_L$, the linear wave, is a solution of the linear Schr\"odinger equation (\ref{S1V}):
 
 \beqa &&{i}\hbar\cdot  \frac{\partial \Psi_L(t,{\bf x})}{\partial t}=
 -\frac{\hbar^2}{2m}\Delta\Psi_L(t,{{\bf x}})+V^{L}(t,{{\bf x}})\Psi_L(t,{{\bf x}}),\label{S1V}\eeqa

 The factorization ansatz results from the recognition that, due to the fundamental non-linearity of the wave dynamics, a linear partition of the type $\Psi(t,{{\bf x}})=\Psi_L(t,{{\bf x}})+ \phi_{NL}(t,{{\bf x}})$ is irrelevant. From this point of view the factorization ansatz incorporates non-linearity from the beginning.
 
 Originally, this ansatz has been introduced by us \cite{new} in order to describe the phenomenology of ``walkers'' (also called bouncing oil droplets\footnote{These are macroscopic objects that exhibit certain quantum-like features. In particular their average trajectories seemingly obey a pseudo dB-B dynamics. In ref.\cite{new}, we simulated the properties of bouncing oil droplets by representing through $\Psi_L(t,{{\bf x}})$ the medium (oil bath) on which droplets propagate and through $\phi_{NL}(t,{{\bf x}})$ the droplets themselves. We derived in that paper an expression for the pseudo-gravitational interaction between two droplets, assuming from the beginning that dB-B guidance equation (\ref{dBB}) was satisfied. In the present paper, we focus on the single particle (droplet) case. We aim here at deriving the dB-B guidance equation from the ansatz (\ref{ansatz}).}). In the case of droplets, our basic motivation for imposing the factorization ansatz is that walkers prepared at different positions and represented by $\phi^i_{NL}(t,{{\bf x}}) (i=1,2...)$ always see the same bath (environment) represented by $\Psi_L(t,{{\bf x}})$. In the same paper \cite{new}, we extended this idea to arbitrary quantum systems, for instance to elementary particles and/or atoms molecules and so on. 
 
In our (wave monist) approach, $\Psi(t,{{\bf x}})$ is assumed to represent the full reality of the quantum system. According to our factorization ansatz $\Psi(t,{{\bf x}})$ can be split into the particle represented by $\phi_{NL}(t,{{\bf x}})$ and in the ``linear'' wave represented by $\Psi_L(t,{{\bf x}})$. The spatial size of the particle being assumed to be extremely small (every particle is a tiny black hole in our approach \cite{new}), the experimenter has supposedly no direct access to/control on their location. In our view, identical experimental preparations however result in the same value for $\Psi_L(t,{{\bf x}})$, which, considered so, does not represent the full information about the system but the information accessible to and controllable by the experimentalist.  As we shall show, $\Psi_L$ can be interpreted as a pilot-wave, while  $\phi_{NL}$ behaves as a solitary wave moving, in good approximation, in accordance with a generalized dB-B guidance equation (\ref{dBB}). Roughly summarized, our main results are the following: 
   
   {\bf Property 1}
   
   whenever $\phi_{NL}$ remains peaked throughout time in a sufficiently small region, its barycentre (from now on denoted ${\bf x_0}$) obeys, in good approximation, the generalized guidance equation 
   
   \beqa{\bf v}_{drift}&=&{\hbar \over m}{{\bf \bigtriangledown}}\varphi_L({{\bf x_0}}(t),t)+{<\phi_{NL}| {\hbar\over i m}{\bf \bigtriangledown}|\phi_{NL}>\over <\phi_{NL}|\phi_{NL}>} \nonumber \\&=&{\bf v}_{dB-B}+{\bf v}_{int.},\label{drift}\eeqa
   
   which contains the well-known Madelung-de Broglie-Bohm contribution (${\bf v}_{dB-B}={\hbar \over m}{{\bf \bigtriangledown}}\varphi_L({{\bf x_0}}(t),t)$) plus a new contribution due to the internal structure of the soliton (${\bf v}_{int.}={<\phi_{NL}| {\hbar\over i m}{\bf \bigtriangledown}|\phi_{NL}>\over <\phi_{NL}|\phi_{NL}>}$). 
   %If we repeat this scheme for Dirac's electron,  $\phi_{NL}$ obeys exactly the dB-B guidance equation associated to Dirac's equation \cite{Takaba,Holland}. 
   
  {\bf Property 2} 
  
 Denoting $\Psi_L= A_Le^{i\varphi_L}$, where $A_L$ is a real amplitude and $\varphi_L$ a real phase, and defining $\phi'_{NL}\equiv\phi_{NL}/A_L$ we find that  the L$_2$ norm of $\phi'_{NL}$ remains constant throughout time in good approximation, while 
 %$\phi'_{NL}$ is a solitary wave that we expect to be close in shape to the ground state of (\ref{NS},\ref{Choquard}).
the solution $\Psi$ obeys
  
  \beqa \Psi(t,{\bf x})\approx e^{i\varphi_L(t,{\bf x})}\phi'_{NL}(t,{\bf x}),\eeqa 
  which confirms indirectly de Broglie's double solution program according to which the linear wave $\Psi_L$  does not represent the particle. Here $\Psi$ represents the particle and it is essentially equal to the product of the soliton $\phi'_{NL}$ (which plays the role of de Broglie's second solution here) with $e^{i\varphi_L}$, the phase of the ``pilot-wave'' $\Psi_L$. $A_L$, the amplitude of the linear wave function, plays here the role of an auxiliary computation tool.

     What we shall not prove rigorously in the present paper is the stability of the soliton, in the sense that we assume from the beginning that the soliton remains peaked throughout time, due to the self-focusing nature of the non-linear self-interaction to which it is submitted. However, if this stability condition is satisfied, then the properties 1 and 2 can be established by lengthy but straightforward computations that we shall detail in the core of the paper. Actually, we independently established by numerical simulations, in a special case (homogeneous self-gravitating sphere in the case where the extent of the ground state is quite smaller than the radius of the sphere),  that stability is de facto guaranteed while properties 1 and 2 are satisfied in very good approximation.

The paper is structured as follows.

In section \ref{notfree} we derive the aforementioned properties 1 and 2, concerning the velocity of the barycentre of $\phi_{NL}$ and its scaling. In section \ref{numeric} we present the confirmations of properties 1 and 2 obtained from numerical simulations. Those simulations also confirm the stability of the soliton, provided the self-focusing is strong enough. Moreover they show that in the classical, non-relativistic regime, the trajectories of the solitons are classical and do not obey the de Broglie guidance law. This property can be explained in terms of a generalized Ehrenfest's theorem. This leads us to formulate in section \ref{conjecture} a conjecture (main conjecture) according to which de Broglie guidance equation is valid ``in average'' due to the presence of an external stochastic field acting at the level of individual velocities. The last section is devoted to discussions and conclusions. In appendix (section \ref{Diracsect}), we attempt to generalize the previous results to Dirac's equation.

\section{\label{notfree}Factorisability ansatz, solitary waves and generalized dB-B guidance.}
We now assume that at the same time an external potential acts on the particle together with a non-linear self-focusing potential of a gravitational nature. We simultaneously impose the  factorisability ansatz. Therefore equations (\ref{nonfreeNL},\ref{ansatz}) are valid.
Substituting (\ref{ansatz}) in (\ref{nonfreeNL}) we get
\beqa &&{i}\hbar\cdot (( \frac{\partial \Psi_L(t,{\bf x})}{\partial t})\phi_{NL}(t,{\bf x})+\Psi_L(t,{{\bf x})}\cdot (\frac{\partial \phi_{NL}(t,{{\bf x}})}{\partial t}))=\nonumber\\ \label{ansatz2}
&&-\frac{\hbar^2}{2m}\Delta\Psi_L(t,{{\bf x}})\cdot \phi_{NL}(t,{{\bf x}})\nonumber \\
&&-\frac{\hbar^2}{2m}(2{\bf \bigtriangledown}   \Psi_L(t,{{\bf x}}) \cdot  {\bf \bigtriangledown}   \phi_{NL}(t,{{\bf x}})     +\Psi_L(t,{{\bf x}})\cdot \Delta\phi_{NL}(t,{{\bf x}}))\nonumber \\&+&V^L\Psi(t,{{\bf x}})+V^{NL}(\Psi)\Psi(t,{{\bf x}}),\eeqa
that, making use of the identity 

${\bf \bigtriangledown}   \Psi_L(t,{{\bf x}})=({\bf \bigtriangledown}   A_L(t,{{\bf x}}))e^{i\varphi_L(t,{{\bf x}})}+\Psi_L(t,{{\bf x}})i{\bf \bigtriangledown}\varphi_L(t,{{\bf x}})$, we replace by a system of two equations\footnote{This replacement is not one to one in the sense that there could exist solutions of equation (\ref{ansatz2}) that do not fulfill the system (\ref{S1V},\ref{S2}). In any case, we focus on a particular class of solutions here.}:

-the linear Schr\"odinger equation

\beqa &&{i}\hbar\cdot  \frac{\partial \Psi_L(t,{\bf x})}{\partial t}=
 -\frac{\hbar^2}{2m}\Delta\Psi_L(t,{{\bf x}})+V^{L}(t,{{\bf x}})\Psi_L(t,{{\bf x}}),\nonumber\eeqa
 
  and the non-linear equation
  \beqa &&{i}\hbar\cdot  \frac{\partial \phi_{NL}(t,{{\bf x}})}{\partial t}=\nonumber
-\frac{\hbar^2}{2m}\cdot \Delta\phi_{NL}(t,{{\bf x}}) \\& -&\frac{\hbar^2}{m}\cdot (i{\bf \bigtriangledown}  \varphi_L(t,{{\bf x}}) \cdot {\bf \bigtriangledown} \phi_{NL}(t,{{\bf x}})+\frac{{\bf \bigtriangledown}  A_L(t,{{\bf x}})}{A_L(t,{{\bf x}})} \cdot   {\bf \bigtriangledown}  \phi_{NL}(t,{{\bf x}}))\nonumber\\ &+&V^{NL}(\Psi)\phi_{NL}(t,{{\bf x}})\label{S2} \eeqa

   In order to solve the system of equations (\ref{S1V},\ref{S2}), it is worth noting that while the L$_2$ norm of the linear wave $ \Psi_L$ is preserved throughout time, because (\ref{S1V}) is unitary, this is no longer true in the case of the non-linear wave $\phi_{NL}$, because the terms mixing $\Psi_{L}$ and $\phi_{NL}$ are not hermitian.

   By a straightforward but lengthy computation that we reproduce integrally in appendix, we established the following result: 
   
   The change of norm of $\phi_{NL}$ obeys
   \beqa{d <\phi_{NL}|\phi_{NL}>\over dt}\approx \frac{\hbar}{m}{ \Delta}  \varphi_L(t,{{\bf x_0}}) \cdot <\phi_{NL}    |\phi_{NL}>\nonumber\\ -2\frac{{\bf \bigtriangledown}  A_L(t,{{\bf x_0}})}{A_L(t,{{\bf x_0}})}\cdot 
\int d^3{\bf x} (\phi_{NL}(t,{{\bf x}}))^*\frac{\hbar\bf \bigtriangledown} {mi } \cdot     \phi_{NL}(t,{{\bf x}}).\label{normchangebis}\eeqa
\subsection{Property 1}
Let us now consider the barycentre ${\bf x_0}$ of the soliton: ${\bf x_0}\equiv {  <\phi_{NL}| {\bf x}|\phi_{NL}>\over <\phi_{NL}|\phi_{NL}>} $ in order to estimate its velocity ${\bf v}_{drift}$:

\beqa {\bf v}_{drift}\equiv{d({  <\phi_{NL}| {\bf x}|\phi_{NL}>\over <\phi_{NL}|\phi_{NL}>} ) \over dt       }\label{defdrift}\eeqa

 For instance, if we consider its $z$ component:

$z_0={<\phi_{NL}|z|\phi_{NL}>\over <\phi_{NL}|\phi_{NL}>}$ and 

${d z_0\over dt}={1\over <\phi_{NL}|\phi_{NL}>}{d<\phi_{NL}|z|\phi_{NL}>\over dt}-{z_0\over <\phi_{NL}|\phi_{NL}>}{d<\phi_{NL}|\phi_{NL}>\over dt},$

so that we find (making use of (\ref{normchangebis}) as well as of results in appendix, section \ref{appendicit})

\beqa &&{d z_0\over dt}= \nonumber {1\over <\phi_{NL}|\phi_{NL}>} \int d^3{\bf x} (\phi_{NL}(t,{{\bf x}}))^*\frac{\hbar\bf \bigtriangledown_z} {mi } \cdot     \phi_{NL}(t,{{\bf x}})
\\ &+&{1\over <\phi_{NL}|\phi_{NL}>} \int d^3{\bf x} (\phi_{NL}(t,{{\bf x}}))^*(\frac{\hbar\bf \bigtriangledown_z} {m } \cdot   \varphi_{L}(t,{{\bf x}}))  \phi_{NL}(t,{{\bf x}})\nonumber 
\\&+&  \nonumber {1\over <\phi_{NL}|\phi_{NL}>}   <\phi_{NL}|(\frac{\hbar}{m}{ \Delta}  \varphi_L(t,{{\bf x}}) ) \cdot z|\phi_{NL}>   
 \\ &+&   {1\over <\phi_{NL}|\phi_{NL}>}  \frac{\hbar}{im}\int d^3{\bf x} \frac{{\bf \bigtriangledown}  A_L(t,{{\bf x}})}{A_L(t,{{\bf x}})} \cdot   {\bf \bigtriangledown}  (\phi_{NL}(t,{{\bf x}}))^*\cdot z \cdot \phi_{NL}(t,{{\bf x}})
\nonumber 
\\ &-& {1\over <\phi_{NL}|\phi_{NL}>}  \frac{\hbar}{im}\int d^3{\bf x}  (\phi_{NL}(t,{{\bf x}}))^*\cdot z \cdot \frac{{\bf \bigtriangledown}  A_L(t,{{\bf x}})}{A_L(t,{{\bf x}})} \cdot   {\bf \bigtriangledown}  \phi_{NL}(t,{{\bf x}}))  \nonumber 
\\&-&{z_0\over <\phi_{NL}|\phi_{NL}>}\cdot (\frac{\hbar}{m}){ \Delta}  \varphi_L(t,{{\bf x_0}}) \cdot<\phi_{NL}    |\phi_{NL}>\label{trip}
 \\ \nonumber&+&2{z_0\over <\phi_{NL}|\phi_{NL}>}\frac{{\bf \bigtriangledown}  A_L(t,{{\bf x_0}})}{A_L(t,{{\bf x_0}})}\cdot \int d^3{\bf x} (\phi_{NL}(t,{{\bf x}}))^*\frac{\hbar\bf \bigtriangledown} {mi } \cdot     \phi_{NL}(t,{{\bf x}}))\eeqa

Now, $\frac{\hbar}{im}\int d^3{\bf x}(\phi_{NL}(t,{{\bf x}}))^*\cdot z \cdot \frac{{\bf \bigtriangledown}  A_L(t,{{\bf x}})}{A_L(t,{{\bf x}})} \cdot   {\bf \bigtriangledown}  \phi_{NL}(t,{{\bf x}})$ 

$\approx z_0 \frac{{\bf \bigtriangledown}  A_L(t,{{\bf x_0}})}{A_L(t,{{\bf x_0}})}\int d^3{\bf x} (\phi_{NL}(t,{{\bf x}}))^*\frac{\hbar\bf \bigtriangledown} {mi } \cdot     \phi_{NL}(t,{{\bf x}})) $, 

while

$<\phi_{NL}|(\frac{\hbar}{m}){ \Delta}  \varphi_L(t,{{\bf x}})  \cdot z|\phi_{NL}>$ $\approx z_0\cdot (\frac{\hbar}{m}){ \Delta}  \varphi_L(t,{{\bf x_0}}) \cdot  <\phi_{NL}|\phi_{NL}>$ and so on so that finally only the two first lines of (\ref{trip}) survive. We get thus the generalized dB-B guidance equation (\ref{drift}), which constitutes the 

 {\bf Property 1}:

 \beqa{\bf v}_{drift}&=&{\hbar \over m}{{\bf \bigtriangledown}}\varphi_L({{\bf x_0}}(t),t)+{<\phi_{NL}| {\hbar\over i m}{\bf \bigtriangledown}|\phi_{NL}>\over <\phi_{NL}|\phi_{NL}>} \nonumber \\&=&{\bf v}_{dB-B}+{\bf v}_{int.}.\nonumber\eeqa
${\bf v}_{drift}$ contains the de Broglie-Bohm velocity
\beqa {\bf v}_{dB-B}\equiv{\hbar \over m}{{\bf \bigtriangledown}}\varphi_L({{\bf x_0}}(t),t),\label{dBB},\eeqa

 and the internal velocity
 \beqa{\bf v}_{int.}\equiv{<\phi_{NL}| {\hbar\over i m}{\bf \bigtriangledown}|\phi_{NL}>\over <\phi_{NL}|\phi_{NL}>}.\label{int.}\eeqa
(\ref{dBB}) is nothing else than de Broglie-Bohm's guidance equation \cite{Holland}, while ${\bf v}_{int.}$ can be considered as a contribution to the average velocity originating from the internal structure of the soliton. Both contributions to the drift are evaluated at the barycentre of the soliton, ${\bf x_0}$. 
\subsection{Property 2}
Let us now consider the change of norm of $\phi_{NL}$.

To do so, we introduce the total time derivative of $A_L$ (${d A_L\over dt}={\partial A_L\over \partial t}+{\bf v}_{drift}\cdot {{\bf \bigtriangledown}} A_L$)
 where ${\bf v}_{drift}={{d  <\phi_{NL}| {\bf x}|\phi_{NL}>  \over <\phi_{NL}|\phi_{NL}>       }\over dt}$ obeys the generalized dB-B guidance equation (\ref{drift}).

By a direct computation, we find

\beqa {{d A_L\over dt}\over A_L}={1\over A_L}({\partial A_L\over \partial t}+{{\bf \bigtriangledown}} A_L\cdot\frac{\hbar\bf \bigtriangledown} {m } \cdot   \varphi_{L}(t,{{\bf x_0}}))+\nonumber\\ {1\over A_L}{{\bf \bigtriangledown}} A_L\cdot{1\over <\phi_{NL}|\phi_{NL}>} \int d^3{\bf x} (\phi_{NL}(t,{{\bf x}}))^*\frac{\hbar\bf \bigtriangledown} {mi } \cdot     \phi_{NL}(t,{{\bf x}})\label{step}\eeqa

Making use of the conservation equation of the linear Schr\"odinger equation ${\partial A^2_L\over \partial t}=-div(A^2\frac{\hbar\bf \bigtriangledown} {m } \cdot   \varphi_{L}(t,{{\bf x_0}}))$ we find 

${1\over A_L}({\partial A_L\over \partial t}+{{\bf \bigtriangledown}} A_L\cdot\frac{\hbar\bf \bigtriangledown} {m } \cdot   \varphi_{L}(t,{{\bf x_0}}))={-1\over 2}div(\frac{\hbar\bf \bigtriangledown} {m } \cdot   \varphi_{L}(t,{{\bf x_0}}))$ and we can rewrite (\ref{step}) as follows:

\beqa {{d A_L\over dt}\over A_L}={-1\over 2}\frac{\hbar}{m}{ \Delta}  \varphi_L(t,{{\bf x_0}})\nonumber\\+{{{\bf \bigtriangledown}} A_L\over A_L}\cdot{1\over <\phi_{NL}|\phi_{NL}>} \int d^3{\bf x} (\phi_{NL}(t,{{\bf x}}))^*\frac{\hbar\bf \bigtriangledown} {mi } \cdot     \phi_{NL}(t,{{\bf x}})\label{step2}\eeqa

Making use of (\ref{normchangebis}) (derived in appendix, section \ref{appendicit}), we obtain at the end

${{d A_L\over dt}\over A_L}={-1\over 2}{1\over <\phi_{NL}|\phi_{NL}>}{d <\phi_{NL}|\phi_{NL}>\over dt}$ so that, finally,

                            \begin{equation}{{d<\phi_{NL}|\phi_{NL}>\over dt}\over <\phi_{NL}|\phi_{NL}>}= -2{{d A_L\over dt}\over A_L}.\label{scalcons}\end{equation}
    
    %${\bf v}_{dB-B}$ appears in (\ref{drift}) as a consequence of the coupling to $\varphi_L$ through the term $-i{\hbar^2\over m}{\bf \bigtriangledown}  \varphi_L(t,{{\bf x}}) \cdot {\bf \bigtriangledown} \phi_{NL}(t,{{\bf x}})$ present in the Hamiltonian of equation (\ref{S2}), that we shall from now on call the guidance potential. The coupling to $A_L$, that is to say, the term $-{\hbar^2\over m}{{\bf \bigtriangledown}  A_L(t,{{\bf x}})\over A_L} \cdot {\bf \bigtriangledown} \phi_{NL}(t,{{\bf x}})$ present in the Hamiltonian of equation (\ref{S2}), that we shall from now on call the $A_L-\phi_{NL}$ potential, does not contribute directly to the drift velocity but as we showed it contributes to the scaling. It also contributes indirectly to the drift velocity because it influences the shape of $\phi_{NL}$, and thus also influences ${\bf v}_{int.}$. 
    
    From the constraint (\ref{scalcons}) we infer the
    
     {\bf Property 2}
     
       \beqa{<\phi_{NL}|\phi_{NL}>(t)\over <\phi_{NL}|\phi_{NL}>(t=0)}={A^2_L(t=0)\over A^2_L(t)},\eeqa where we evaluate $A^2_L(t)$ at the barycentre of $\phi_{NL}$, which moves according to the generalized dB-B guidance equation (\ref{drift}).
       Let us rescale  $\phi_{NL}(t,{{\bf x}})$ by defining $\phi'_{NL}$ through $\phi_{NL}(t,{{\bf x}})\equiv \phi'_{NL}(t,{{\bf x}})/A_L$; 
       we can thus predict in general that, if it exists and remains peaked during its evolution, the solution of  (\ref{S2}) has the form 
  \begin{equation}\Psi(x,y,z,t) \approx \phi'_{NL}({\bf x},t)e^{i\varphi_L({\bf x},t)},\end{equation}  where $\phi'_{NL}({\bf x},t)$ is centred in $  {\bf x}_0(t=0)+\int_0^t dt {\bf v}_{drift}$ and is of constant L$_2$ norm.     
    Now, $V^{NL}(\Psi)=V^{NL}(\phi'_{NL}(t,{{\bf x}}))$, in virtue of (\ref{Coulomb}) (appendix) and (\ref{S2}) can then be cast in the form
 \beqa &&{i}\hbar\cdot  \frac{\partial ( \phi'_{NL}(t,{{\bf x}})/A_L(t,{{\bf x_0}}))}{\partial t}\nonumber
 =\\ \nonumber
 &&-\frac{\hbar^2}{2m}\cdot \Delta ( \phi'_{NL}(t,{{\bf x}})/A_L(t,{{\bf x_0}}))+V^{NL}(\phi'_{NL})( \phi'_{NL}(t,{{\bf x}})/A_L(t,{{\bf x_0}}))\nonumber\\ &&-\frac{\hbar^2}{m}\cdot i{\bf \bigtriangledown}  \varphi_L(t,{{\bf x}}) \cdot {\bf \bigtriangledown} (\phi'_{NL}(t,{{\bf x}})/A_L(t,{{\bf x_0}})) \nonumber\\ &&-\frac{\hbar^2}{m}\cdot\frac{{\bf \bigtriangledown}  A_L(t,{{\bf x}})}{A_L(t,{{\bf x}})} \cdot   {\bf \bigtriangledown}  ( \phi'_{NL}(t,{{\bf x}})/A_L(t,{{\bf x_0}})).
 \label{S3}\eeqa
 
 In order to say more about $\phi'_{NL}({\bf x},t)$ we must solve equation (\ref{S3}) which is a complicated problem, beyond the scope of our paper.

{\bf Remark.}

   As has been noted in \cite{old}, when there is no external potential ($V^L$), which means that the particle is only submitted to its self-interaction, %equation (\ref{soliton}) describes an exact solution provided the linear wave is a plane wave moving at velocity $v_{dB-B}$. Then%
    we find the exact solution $\phi^0_{NL}({\bf x}-{\bf v}\cdot t)e^{-i((E_0+\hbar \omega)\cdot t-\hbar {\bf k}\cdot {\bf x})/\hbar}$. This is a plane wave moving at velocity $v_{dB-B}$.  As already noted by Fargue \cite{Fargue} many years ago, such solutions exist for a large class of different non-linearities (e.g. a non-linearity proportional to $|\Psi|^2$), and they all agree with the principle of phase concordance proposed by de Broglie in 1927. Actually, this class of solution is well-known and it can be generated from the static solution $\phi^0_{NL}({\bf x})e^{-iE_0 t/\hbar}$ by a Galilean boost. In the present case, this property is seen to be a direct consequence of the Galilean invariance of equation  (\ref{NS}).
   
% In first approximation, \beqa\Psi(x,y,z,t)\approx e^{-iE_0 t/\hbar}\phi^0_{NL}({\bf x}- \int_0^t dt {\bf v}_{dB-B})e^{i\varphi_L({\bf x},t)}.\label{soliton}\eeqa 

%This is a crude approximation however, and all 

% It is worth repeating however that if we estimate ${\bf v}_{int.}$ at the first order of linear perturbation theory,  that is to say if we approximate $\phi'_{NL}({\bf x},t)e^{i\varphi_L({\bf x},t)}$ by the solitary wave defined in (\ref{soliton}), ${\bf v}_{int.}$ = 0, because ${\bf v}_{int.}$ is proportional to the average velocity of the ``free'' soliton, solution of (\ref{Choquard}) evaluated in the frame where it is at rest. 
  
   % What remains to do in order to find better approximations of the exact solution of  (\ref{S2}) is to resort to perturbative methods. This is appropriate, having in mind, that the L$_2$ norm of $\phi'_{NL}$ does not vary and that the non-linear potential ``seen'' by $\phi'_{NL}$ does not rescale,  in virtue of (\ref{Coulomb}), which opens the door to a perturbative approach \cite{new}. 
%One can actually show that the first order corrections estimated by averaging the error made in imposing an approximate solution of the type (\ref{soliton}) in (\ref{nonfreeNL}) goes to zero when the size of the soliton $\phi^0_{NL}$ decreases \cite{prepa}.

\section{Main conjecture.}
\subsection{Numerical simulations.\label{numeric}}
{\bf Normalisation and Born rule.}

 One could object that in order to fit to the constraints required by our model, in particular in order to ensure that the size of the soliton is quite smaller than the size of the linear wave, the coupling constant ought to be huge. However this is not true. We are free to rescale the solitonic solutions of (\ref{Choquard}) without being constrained by the normalisation to unity of the wave function, which is a condition imposed by the Born rule. In our case, the Born rule is not postulated to begin with, it should rather be derived from the so-called equilibrium condition according to which the statistical distribution of the positions of the particles asymptotically converges in time to the Born distribution in $|\Psi_L|^2$. In our eyes, the equilibrium condition ought to be derived from the generalized dB-B dynamics as is done in e.g. classical chaos theory\footnote{There exist serious attempts to derive the onset of quantum equilibrium from the de Broglie-Bohm mechanics \cite{tony,grec,colin,ColinStruyve,grec2,ab}, but this is a deep and complex problem, reminiscent of the H-theorem of Boltzmann, which opens the door to a Pandora box that we do not wish to open here (see papers of Colin, Durt and Willox, of Drezet and of Efthymiopoulos, same issue).}. It is well-known for instance from the study of deterministic chaotic systems that the sensitivity to initial conditions is an essential ingredient for generating stochasticity and impredictability. This ingredient is present in the dB-B  \cite{butterfly} dynamics too.
 
 {\bf Self-gravitating nanosphere in an external one-dimensional harmonic trap.}
 
 We shall now briefly mention some results obtained by us regarding the properties of a self-gravitating nanosphere placed in an harmonic trap, relaxing the constraint on the normalisation of the wave function. For convenience, we treat the system as an homogeneous sphere, in the limit where the extent of the wave function of its center of mass is quite smaller than its radius. In this case, $V^L=k^{ext.} x^2/2$ while (see \cite{diosi84,Chen,CDW} and discussion of footnote \ref{coco}).
  \beqa V^{NL}(x)= {GM^2 \over R}\int {d} x' |\Psi(t,x')|^2(-\frac{6}{5}+\frac{1}{2}(\frac{|x-x'|}{R})^2+{\cal O}((\frac{|x-x'|}{R})^3))\nonumber \\ \approx N^2({-6GM^2 \over 5R}+{GM^2 \over 2R^3}(<x^2>-<x>^2))+{GM^2 \over 2R^3}N^2(x-<x>)^2,\eeqa
 
 with $N^2=\int {d} x' |\Psi(t,x')|^2$,  
 
 $<x>=\int {d} x' |\Psi(t,x')|^2 x'/N^2$ and $<x^2>=\int {d} x' |\Psi(t,x')|^2 (x')^2/N^2$.
 
 Denoting $k^{self}={GM^2 \over 2R^3}N^2$ and reexpressing the Hamiltonian up to irrelevant position-independent factors, we get the following evolution equation:
 
 \begin{equation}
{i}\hbar\frac{\partial\Psi(t,x)}{\partial t}=-({\hbar^2\over 2m})\frac{\partial^2\Psi(t,x)}{\partial x^2}
+({k^{ext.}\over 2} x^2+{k^{self}\over 2}(x-<x>)^2)\Psi(t,x).\label{Carlos}
\end{equation}

 {\bf Numerical simulations.}
 
The main advantage of this simple model is that it is gaussian: initially gaussian states remain gaussian throughout their temporal evolution under (\ref{Carlos}). Therefore we were able to numerically solve it with high accuracy, which would be impossible with for instance the single particle N-S equation (\ref{NS}) \cite{ducan}. 

$N^2$, the square of the $L_2$-norm of $\Psi$ ($N^2=\int {d} x' |\Psi(t,x')|^2$) is a conserved quantity under the evolution (\ref{Carlos}); it is however a free parameter in our approach (in agreement with the discussion of the previous paragraph), which means that we are essentially free to choose the value of the effective self-gravitating constant $k^{self}$. Varying this free parameter, we studied the behavior of gaussian solutions (denoted $\Psi^{NL}$) of the non-linear equation (\ref{Carlos}). We also decomposed the initial gaussian states $\Psi^{NL}(t,x=0)$ according to the ansatz (\ref{ansatz}): 

$\Psi^{NL}(x,t=0)=\Psi^{L}(x,t=0)\cdot \phi_{NL}(x-<x>_0,t=0)$, 

where $\phi_{NL}(x-<x>_0,t=0)$ is a gaussian real state, of which the shape is close to the shape of the ground state $\phi^0_{NL}(x)$ of the ``free'' equation ${i}\hbar\frac{\partial\Psi(t,x)}{\partial t}=-({\hbar^2\over 2m})\frac{\partial^2\Psi(t,x)}{\partial x^2}
+({k^{self}\over 2}(x-<x>)^2)\Psi(t,x),$ which is \cite{CDW} a real gaussian state centered in  $<x>_0$ of extent $\sqrt{\hbar \over \sqrt{k^{self}\cdot m}}$. $\Psi^{L}(x,t)$ and $\Psi^{NL}(x,t)$ were obtained by integrating respectively the linear equation  ${i}\hbar\frac{\partial}{\partial t}\Psi^L(t,x)=-({\hbar^2\over 2m})\frac{\partial^2}{\partial x^2}\Psi^L(t,x)
+({k^{ext.}\over 2} x^2)\Psi^L(t,x)\label{linear}$ and equation (\ref{Carlos}). $<x>_0$ was chosen freely, excepted that we imposed that $\Psi^L(x,t=0)=\Psi^{NL}(x,t=0)/\phi^{NL}(x,t=0)$ was a normalisable (gaussian) state. This allowed us to estimate the solitonic wave function $\phi^{NL}(x,t)=\Psi^{NL}(x,t)/\Psi^L(x,t)$ at all times. Our first result concerns stability: our criterion for stability is that $\phi^{NL}(x,t)=\Psi^{NL}(x,t)/\Psi^L(x,t)$ is a normalisable (gaussian) state. This means that the soliton remains peaked during time. We noticed that when $k^{self}/k^{ext.}$ was high enough (larger than 10 was obviously sufficient), stability was ensured. Not surprisingly, in this case, the self-gravitation forces the soliton to oscillate around its center, without ever escaping to the self-focusing gravitational potential ${k^{self}\over 2}(x-<x>)^2$. 

\begin{figure}[h]\vskip.5cm
\begin{center}
\resizebox{10cm}{!}{\includegraphics[]{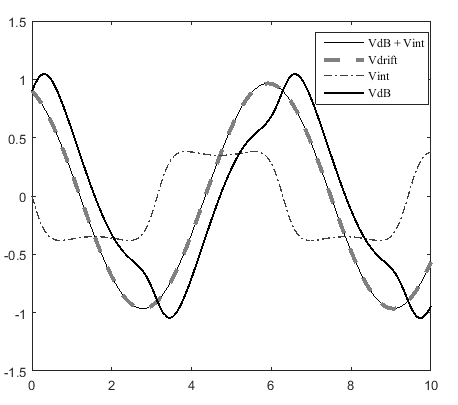}}\end{center}
\caption{Plot of $v_{dBB}, v_{int.}, v_{dBB}+ v_{int.}$ and $v_{drift}={d\over dt}(<x_{NL}>)$ with $<x_{NL}>=\int {d}^3 x' |\phi^{NL}(t,x')|^2 x'/N_{NL}^2$) in function of time; space and time were rescaled and are of the order of unity.}\label{Momo}
\end{figure}

 The results of our numerical study are encapsulated in the figure \ref{Momo} \cite{Momo}  which represents the velocity $v_{drift}\equiv{d\over dt}(<x_{NL}>)$ (with $<x_{NL}>=\int {d}^3 x' |\phi^{NL}(t,x')|^2 x'/N_{NL}^2$) of the centre of the soliton $\phi^{NL}(x,t)$, in function of time. 

To derive the plot \ref{Momo}, we imposed the following constraints: 
 
 $k^{self}/k^{ext.}= 10^3$,
   $(<x_{NL}^2>-<x_{NL}>^2)_{t=0}= \sqrt{\hbar \over \sqrt{(k^{ext}+k^{self})\cdot m}}$,
   
    and $(<x_{NL}^2>-<x_{NL}>^2)_{t=0}=10^{-3}(<x_{L}^2>-<x_{L}>^2)_{t=0}$.

 We observed that 

(i)  In the limit where $k^{self}/k^{ext.}$ is large enough, and provided at time $t=0$ the extent of the soliton $(<x_{NL}^2>-<x_{NL}>^2)_{t=0}$ is comparable to the extent of the self-focused ground state ($\sqrt{\hbar \over \sqrt{k^{self}\cdot m}}$), and also quite smaller than the extent of the linear wave ($(<x_{NL}^2>-<x_{NL}>^2)_{t=0}$ $<<$ ($<x_{L}^2>-<x_{L}>^2)_{t=0}$), $\phi^{NL}(x,t)$ remains strongly peaked and oscillates around its central value throughout time.

(ii) In good approximation (the quality of the approximation being an increasing function of the ratio $k^{self}/k^{ext.}$), property 1 is confirmed by our numerical estimates. Indeed, as can be seen in  figure \ref{Momo}, $v_{drift}$ is indistinguishable from $v_{dBB}+ v_{int.}$.

(iii) In these conditions, property 2 is automatically satisfied.

(iv) We also noted that in good approximation (the quality of the approximation being an increasing function of the ratio $k^{self}/k^{ext.}$) the soliton $\phi^{NL}(x,t)$, as well as  $\Psi^{NL}(x,t)$, follow classical trajectories.

\subsection{Main conjecture.\label{conjecture}}
The numerical results presented in the previous chapter (in particular observation (iv)) brought us to demonstrate the

{\bf Generalized Ehrenfest theorem.} 

To do so, let us consider the evolution equation 
\begin{equation}
{i}\hbar\frac{\partial\Psi(t,{{\bf x}})}{\partial t}=(-\hbar^2\frac{\Delta}{2m}+V^{L}(t,{{\bf x}})
+V^{NL}(t,{{\bf x}}))\Psi(t,{{\bf x}})\label{NSgen},\end{equation} where $V^{NL}t,{{\bf x}})=-Gm^2\int {d}^3 x' |\Psi(t,{{\bf x}'})|^2F(|{\bf x} -{\bf x'}|)$, with $F$ a real function of $|{\bf x} -{\bf x'}|$.
Then the centre of its solution, denoted $<x>$ ($<x>=\int {d}^3 x |\Psi(t,x)|^2 x/N^2$ with $N^2=\int {d}^3 x' |\Psi(t,x')|^2$) obeys the laws of classical dynamics in the limit where $<x^2>-<x>^2$ goes to zero.

To prove it, it suffices to note that $N$ remains constant throughout time, while 

${d <x>\over d t}=<{i\hbar\over m}{\bf \nabla}>$, and 

${d^2 <x>\over d t^2}={1\over m}<[{\bf \nabla},V^{L}+V^{NL}]>={1\over m}<{\bf \nabla}V^{L}+{\bf \nabla}V^{NL}>.$

Now, $<{\bf \nabla}V^{NL}>$
 
 $=\int {d}^3 x' |\Psi(t,x')|^2\int {d}^3 x |\Psi(t,x)|^2({\bf x} -{\bf x'}) {dF\over du}_{|u=|{\bf x} -{\bf x'}|}=0,$
by symmetry so that 

${d^2 <x>\over d t^2}={1\over m}<{\bf \nabla}V^{L}>$.

In the limit where $<x^2>-<x>^2$ goes to zero, $<{\bf \nabla}V^{L}>\approx {\bf \nabla}V^{L}(<x>)$. Note that when $V^{L}$ is a quadratic function of the position (harmonic oscillator), $<{\bf \nabla}V^{L}>$ is always exactly equal to ${\bf \nabla}V^{L}(<x>).$

At this level, we face a serious problem: our original goal \cite{old} was to derive dB-B dynamics from the non-linear dynamics (\ref{nonfreeNL}) and the factorization ansatz (\ref{ansatz}). However, the trajectory of the soliton departs from dB-B dynamics, due to the presence of the residual contribution ${\bf v}_{int.}={\bf v}_{drift}-{\bf v}_{dB-B}$. Our simulations reveal that, in order to ensure the constraints imposed by Eherenfest's theorem, ${\bf v}_{int.}$ ``conspires'' in order to be equal to ${\bf v}_{classical}-{\bf v}_{dB-B}$. This compensation is fragile however in the sense that ${\bf v}_{classical}-{\bf v}_{dB-B}$ contains the memory of the initial preparation process that eventually took place a long time ago. We shall conjecture here that in the practice some stochasticity is present that will wash out this memory and decorrelate ${\bf v}_{int.}$ from ${\bf v}_{classical}$. Averaging over this stochastic contribution we expect thus (and this is our main conjecture) that $<<{\bf v}_{int.}>>=0$ so that $<<{\bf v}_{drift}>>=<<{\bf v}_{dB-B}>>$, where the bracket $<<,>>$ represents an averaging over this extra-stochastic perturbation of the velocities that we conjecture to be present in nature.

 Note that in the past Bohm, Vigier \cite{BV} and de Broglie \cite{debroglieend} suspected the existence of a stochastic noise superposed to the quantum potential, necessary in their eyes in order to explain the irreversible in time convergence to quantum equilibrium. de Broglie wrote for instance in \cite{debroglieend} (chapter XI: On the necessary introduction of a random element in the double solution theory. The hidden thermostat and the Brownian motion of the particle in its wave) the following sentences

 {\it ...Finally, the particle's motion is the combination of a regular motion defined by the guidance formula, with a random motion of Brownian character... any particle, even isolated, has to be imagined as in continuous ``energetic contact'' with a hidden medium, which constitutes a concealed thermostat. This hypothesis was brought forward some fifteen years ago by Bohm and Vigier \cite{BV}, who named this invisible thermostat the ``subquantum medium''....If a hidden sub-quantum medium is assumed, knowledge of its nature would seem desirable...}

Our conjecture is a re-expression of the subquantum medium hypothesis invoked by Bohm and Vigier. At this level its deep nature is not clear yet: it could be the manifestation of a relativistic effect similar to the zitterbewegung (footnote \ref{footnotezitter} in appendix) or it could be of a fundamental nature \cite{Sonego}. It is out of the scope of the present paper to study these possibilities. It is worth noting however that in the case of bouncing oil droplets\footnote{Of course we do not expect that Planck's constant plays a role in the case of walkers, because instead of Schr\"odinger equation, one should consider the classical wave equation that describes the dynamics of waves at the surface of the oil bath to begin with, which is outside of the scope of the present paper. We expect however that the techniques developed in the present paper are still valid in the classical regime in the case of droplets.  
} \cite{new} such a stochastic disturbance of velocities is always present, due to the periodical forcing imposed to the bath. This explains why dB-B trajectories are never observed directly at the level of droplets but result from the averaging of a large number of trajectories  \cite{fort,fort2,Bush}.

\section{Discussion and Conclusions.}

{\bf Discussion: experimental validation.}

  The overwhelming majority of experiments \cite{smolin,Chen,arxiv,bili,optom2} proposed so far in order to reveal the existence of intrinsic non-linearities at the quantum level (like e.g. the self-gravity interaction) is a priori doomed to fail, for what concerns our model, because their conceivers systematically took for granted that the wave function was normalised to unity\footnote{Even in this case, the generalized Ehrenfest theorem demonstrated in section \ref{conjecture} is valid, which predicts that in the limit of massive enough objects trajectories become classical. This could open the way to new experimental tests for falsifying self-gravity. It also sheds a new light onto the problem of the classical limit (see paper of Matzkin same issue).}. In the same line of thought, if our conjecture is correct, then trajectories of the self-collapsed solitons are  very close to de Broglie-Bohm trajectories. Therefore we expect that our model cannot easily be falsified, to the same extent that the dB-B interpretation leads to the same predictions as the standard quantum theory (through the Born rule).
  
  Even if our model is not relevant, and in the last resort its relevance ought to be confirmed or falsified by experiments, it could appear to be useful as a phenomenological tool. 
  
  For instance it was already applied by us \cite{new} in the past\footnote{In the present paper we essentially studied the feedback of the pilot wave $\Psi_L$ on the soliton or particle $\phi_{NL}$, postulating to begin with that the factorisation ansatz is satisfied and that a self-focusing non-linearity is present, that prohibits the spread of the soliton. In ref. \cite{new}, which is complementary to the present paper we studied the feedback of the dB-B solitons on the pilot wave. We predicted the appearance, when several droplets are present simultaneously, of  an effective gravitational potential and formulated certain predictions that are likely to be tested in the laboratory but this is another story. } to the phenomenology of bouncing oil droplets \cite{fort,fort2,Bush}. It is at this level that we expect that the validity of the approximations and hypotheses of our model could be confirmed or falsified.

  The dynamics outlined here might also be relevant in the field of cold atoms physics \cite{Kaiser} where effective non-linear self-focusing equations properly describe collective excitations of the atomic density \cite{anas}. As far as we know, no de Broglie like trajectory has yet been observed during such experiments. 
  
  In the same order of ideas, it would be highly interesting to investigate whether de Broglie like trajectories are good tools for describing optical and/or rogue waves \cite{akhme,koon}. After all our factorisability ansatz could be applied to classical non-linear wave equations too.

 {\bf Conclusions.}

   %The normalisation of the solitonic solutions considered by us is left entirely unspecified here; we need an argument that fixes the normalisation once for all. It is not clear whether such an argument can be found in existing theories such as the standard model of particles physics. 
   %At the other side, preliminary results tend to indicate that the size of the soliton is imposed by requiring consistency with Newtonian gravity \cite{prepa}. 
   %Related to this, the scaling of the non-linear potential required for localizing the particle is not arbitrary, but exhibits a $|\Psi|^2$ dependence (\ref{Coulomb}), typical of the Newton and/or Coulomb self-interaction \cite{visser}. 
   
     %All these questions are important of course and they deserve to be scrutinized in depth. At this level, our goal is less ambitious. 
     
     We have studied particular solutions of the S-N equation, in presence of an external potential, that behave as peaked solitons, due to the self focusing nature of self-gravitation. We showed that, if they exist and are stable, their shape is close in first approximation to the shape of the self-collapsed ground state, solution of (\ref{Choquard}) and, provided they remain peaked throughout time, they move in accordance with a generalized dB-B dynamics. These predictions were confirmed by numerical simulations. These simulations also showed that in the non-relativistic limit, the generalized dB-B dynamics is equivalent to classical dynamics, which is confirmed by a generalized Ehrenfest theorem. We conjecture however that if we add a stochastic field to the velocity field, dB-B dynamics will be restored. In last resort, this field could be a manifestation of the relativistic zitter bewegung (see appendix). In our approach, the linear wave function plays the role of an auxiliary computation tool (a pilot wave), and does not represent the particle which is represented by the soliton, which fits well into the double solution program of de Broglie. 
     
     We hope that the rather simple models treated in this paper will convince the reader that de Broglie's ideas were maybe not that much surrealistic \cite{surreal} and deprived of consistence. Our results indeed reinforces the dB-B picture according to which the particle non-locally and contextually explores its environment thanks to the nearly immaterial tentacles provided by the solution of the linear Schr\"odinger equation. This picture is not comfortable but it is maybe the price to pay to restore wave monism\footnote{An advantage of wave monism is that, contrary to the Bohmian dynamics formulated in terms of material points, wave monism does not violate the No Singularity Principle formulated by Gouesbet \cite{Gouesbet}. Everything is continuous in our approach, even if very different spatial scales coexist.}. Our conjecture (section \ref{conjecture}) also reemphasises the necessity to invoke the existence of a subquantum medium in order to add to the regular and deterministic dB-B trajectories a random motion of brownian character. The brief incursion in the relativistic domain sketched in appendix suggests that the origin of this brownian motion could be linked to zitterbewegung \cite{Hestenes,Messiah}.  It would be highly interesting to try and reveal the existence of this subquantum medium experimentally, which is seemingly not an impossible task as suggested in ref.\cite{Thrust}.

   Last but not least, our analysis also confirms several prophetic intuitions originally presented by Louis de Broglie during the Solvay conference of 1927 \cite{1927}.

%We cannot resist ending this paper with a quotation from Han Solo in Star Wars 7, The Force Awakens: 

%{\it...It's true. All of it. The Dark Side, the Jedi. They're real....}.

\section*{Acknowledgements}
This work was made possible thanks to the financial support of a grant from the FQXi Foundation (Quantum Rogue Waves as Emerging Quantum Events-FQXI project FQXi-RFP-1506).
It also benefited from the support of grants 21326 ``Non-Linearity and Quantum Mechanics: Quest for a Rogue Wave Mechanics''and 60230 ``Non-Linearity and Quantum Mechanics: Limits of the No-Signaling Condition''  from the John Templeton Foundation. Figure 1 was derived in collaboration with C.E. Lopez-Fortin (master student, Centrale Marseille), and M. Hatifi, Ph.D student, Templeton grant 60230. Sincere thanks to C. Champenois, E. Lassalle, B. Stout, and R. Willox for their comments on a primitive version of the paper.

\vskip 30pt\bibliographystyle{plain}

\section{Appendices.}

\subsection{The Schr\"odinger-Newton equation.\label{didon2}}

One can look for a ``ground-state" solution to \eqref{NS} in the form
\begin{equation}
\psi(\bx, t) = {e}^{-\frac{{i}E_0 t}{\hbar}} \phi^0_{NL}(\bx)~\!,\label{redtoChoq}
\end{equation}

This leads to a stationary equation for $\phi^0_{NL}$
\begin{equation}\frac{-\hbar^2}{2M} \Delta \phi^0_{NL}(\bx) + G M^2\! \int\!d^3y ~\!\frac{\big| \phi^0_{NL}(\by)\big|^2}{|\bx-\by|}~\! \phi^0_{NL}(\bx) = E_0  \phi^0_{NL}(\bx)~\!,\label{Choquard}
\end{equation}
which was studied in astrophysics and is known under the name of the {\em Choquard} equation \cite{Lieb}.  In \cite{Lieb}, Lieb showed that the energy functional 
\begin{equation}
E(\phi) = \frac{\hbar^2}{2 M} \int\!{d}^3x~\big|\nabla\phi(\bx)\big|^2 - \frac{G M^2}{2} \iint\!{d}^3x {d}^3y~\!\frac{~\big|\phi(\by)\big|^2}{|\bx-\by|}~\!\big|\phi(\bx)\big|^2~\!,\label{Choquard-energy}
\end{equation}
is minimized by a unique solution $ \phi^0_{NL}(\bx)$ of the Choquard equation \eqref{Choquard} for a given $L_2$ norm $N( \phi^0_{NL}(\bx))$. However no analytical expression is known for this ground state. Numerical treatments established that this ground state has a quasi-gaussian shape, and that its extent is, in the case  $N( \phi^0_{NL}(\bx))=1$, of the order of ${\hbar^2\over G M^3}$. Then its energy is of the order of $G^2M^5/\hbar^2$. Otherwise, energy scales like $N( \phi^0_{NL}(\bx))^3$ \cite{CDW}. Numerical evidence suggests the existence of a discrete spectrum of values for the energy (\ref{Choquard-energy}) of the bound states solutions of (\ref{Choquard}). They can be written in the form $-\mathrm{e}_n \frac{G^2 M^5}{\hbar^2}$, described in terms of dimensionless constants $\mathrm{e}_n$ as :
\begin{equation}
\mathrm{e}_n = \frac{a}{(n+b)^c}~\!,\quad n= 0, 1, 2, \hdots~\!,\label{Choquard-spectrum}
\end{equation}
for approximated constants \cite{Bernstein}
\begin{equation}
a = 0.096\pm0.01~,\quad b = 0.76\pm 0.01~,\quad c= 2.00\pm 0.01~\!.
\end{equation}
The ``ground state" for the Choquard equation corresponds, at $N=1$, to $n=0$ (see \cite{CDW} and references therein for more details).

In summary, the S-N potential exhibits several interesting features that play an important role in our paper: it is self-focusing and possess a localized ground state of the type $\phi^0_{NL}({\bf x})e^{-iE_0 t/\hbar}$ which behaves as a static bright soliton. The spectrum of negative energy eigenstates solutions of (\ref{Choquard}) is discrete. Moreover this potential scale like a Coulomb and/or Newtonian self-interaction in the sense that
\beqa V^{NL}(\lambda \Psi)=|\lambda|^2V^{NL}(\Psi)\label{Coulomb}\eeqa

\subsection{Change of norm\label{appendicit}}
 Let us denote $H_L$ the linear part of the the full Hamiltonian in (\ref{S2}). It is not hermitian, so that $\sqrt{<\phi_{NL}|\phi_{NL}>}$, the L$_2$ norm of its solution $\phi_{NL}(t,{{\bf x}}))$ is not constant throughout time. The non-linear potentials considered by us preserve the L$_2$ norm however. We can thus evaluate the time derivative of $<\phi_{NL}|\phi_{NL}>$ by direct computation, either integrating by parts, or making use of the formula 
 
 \beqa &&{d <\phi|O|\phi>\over dt}=\nonumber \\&&<\phi|{\partial O\over \partial t}|\phi>+{1\over i\hbar}<\phi|O H_L-H_L^\dagger O|\phi>\nonumber \\&&=<\phi|{\partial O\over \partial t}|\phi>+{1\over i\hbar}( <\phi|[O, Re.H_L]_-|\phi>\\ \nonumber &&+{1\over \hbar}( <\phi|[O, Im.H_L]_+|\phi>),\eeqa
where $O$ is an arbitrary observable, described by a self-adjoint operator,  while $Re.H_L$ and $Im.H_L$, the real and imaginary parts of $H_L$ are self-adjoint operators defined through $2\cdot Re.H_L=H_L+H_L^\dagger$ and $2i\cdot Im.H_L=H_L-H_L^\dagger$. Here, the symbol $[,]_-$ ($[,]_+$) represents the (anti)commutator.

We find by direct computation that 

\beqa Re. (-\frac{\hbar^2}{m}i{\bf \bigtriangledown}  \varphi_L(t,{{\bf x}}) \cdot {\bf \bigtriangledown})\nonumber  \\=
 (-\frac{\hbar^2}{m}i{\bf \bigtriangledown}  \varphi_L(t,{{\bf x}}) \cdot {\bf \bigtriangledown})-(\frac{\hbar^2}{2m}i{ \Delta}  \varphi_L(t,{{\bf x}}) )\eeqa and $Im. (-\frac{\hbar^2}{m}i{\bf \bigtriangledown}  \varphi_L(t,{{\bf x}}) \cdot {\bf \bigtriangledown})=(\frac{\hbar^2}{2m}{ \Delta}  \varphi_L(t,{{\bf x}}) )$. 
 
 Therefore the guidance potential contributes to 
 
 ${d <\phi_{NL}|\phi_{NL}>\over dt}={d <\phi_{NL}|1|\phi_{NL}>\over dt}$ by a quantity 

\beqa <\phi_{NL}|    (\frac{\hbar}{m}{ \Delta}  \varphi_L(t,{{\bf x}}) )    |\phi_{NL}>\approx (\frac{\hbar}{m}{ \Delta}  \varphi_L(t,{{\bf x}}) )<\phi_{NL}    |\phi_{NL}>),\nonumber\eeqa due to the fact that, over the size of the soliton, $\varphi_L(t,{{\bf x}}) $ and its derivatives are supposed to vary so slowly that we can consistently neglect their variation and put them in front of the L$_2$ integral.

The contribution of the $A_L-\phi_{NL}$ coupling to ${d <\phi_{NL}|\phi_{NL}>\over dt}$ is

$\frac{\hbar^2}{m}{1\over i\hbar}\int d^3{\bf x} (\frac{{\bf \bigtriangledown}  A_L(t,{{\bf x}})}{A_L(t,{{\bf x}})} \cdot   {\bf \bigtriangledown}  (\phi_{NL}(t,{{\bf x}}))^*\phi_{NL}(t,{{\bf x}})-  (\phi_{NL}(t,{{\bf x}}))^*\frac{{\bf \bigtriangledown}  A_L(t,{{\bf x}})}{A_L(t,{{\bf x}})} \cdot   {\bf \bigtriangledown}  \phi_{NL}(t,{{\bf x}})).$

We now suppose that we are in right to neglect the variation of $\frac{{\bf \bigtriangledown}  A_L(t,{{\bf x}})}{A_L(t,{{\bf x}})}$ in the integral above and we find, integrating by parts, a contribution $-2\frac{{\bf \bigtriangledown}  A_L(t,{{\bf x_0}})}{A_L(t,{{\bf x_0}})}\cdot 
\int d^3{\bf x} (\phi_{NL}(t,{{\bf x}}))^*\frac{\hbar\bf \bigtriangledown} {mi } \cdot     \phi_{NL}(t,{{\bf x}})$

Putting all these results together, we find that

\beqa{d <\phi_{NL}|\phi_{NL}>\over dt}\approx \frac{\hbar}{m}{ \Delta}  \varphi_L(t,{{\bf x_0}}) \cdot <\phi_{NL}    |\phi_{NL}>\nonumber \\ -2\frac{{\bf \bigtriangledown}  A_L(t,{{\bf x_0}})}{A_L(t,{{\bf x_0}})}\cdot 
\int d^3{\bf x} (\phi_{NL}(t,{{\bf x}}))^*\frac{\hbar\bf \bigtriangledown} {mi } \cdot     \phi_{NL}(t,{{\bf x}}).\label{normchange}\eeqa

\subsection{The Dirac spinor.\label{Diracsect}}
In our view the  S-N equation is a good candidate for fulfilling the de Broglie double solution program of 1927, in the same sense that Schr\"odinger's equation was a good candidate for realizing de Broglie's wave mechanics program of 1925. In the same line of thought, it is natural to look for a non-linear relativistic equation that would be to the S-N equations what are the Klein-Gordon or Dirac equations to the non-relativistic Schr\"odinger equation. 

To do so let us consider the non-linear Dirac equation

\beqa \beta i\hbar \partial_t {\bf \Psi}(t,{\bf x})-\beta{\bf \alpha}c{\hbar\over i}{\bf \bigtriangledown}{\bf \Psi}(t,{\bf x})-(mc^2+V_L){\bf \Psi}(t,{\bf x})-V_{NL}{\bf \Psi}(t,{\bf x})=0.\label{NLD} \eeqa

where $\alpha$ and $\beta$ are the well-known Dirac matrices and \beqa V_{NL}{\bf \Psi}(t,{\bf x})=-Gm^2\int {d}^3 x'(\frac{|\Psi(t,{{\bf x}'})|_4^2}{|{{\bf x} -{\bf x}'}|})\Psi(t,{{\bf x}}).\label{Newton}\eeqa

Here, $|\Psi(t,{{\bf x}'})|_4^2$ represents the local Dirac density: 

$|\Psi(t,{{\bf x}'})|_4^2$=$({\bf \Psi}(t,{\bf x}))^\dagger \cdot_4 {\bf \Psi}(t,{\bf x}))$, where the lower index $_4$ refers to the spinorial in-product.
Let us solve (\ref{NLD}) imposing the ansatz

\beqa {\bf \Psi}=\left(\begin{array}{c}
 \Psi_0(t,{{\bf x}}) \\
 \Psi_1(t,{{\bf x}}) \\
 \Psi_2(t,{{\bf x}}) \\
 \Psi_3(t,{{\bf x}}) \\
\end{array}\right)=\left(\begin{array}{c}
 \Psi^L_0(t,{{\bf x}}) \\
 \Psi^L_1(t,{{\bf x}}) \\
 \Psi^L_2(t,{{\bf x}}) \\
 \Psi^L_3(t,{{\bf x}}) \\
\end{array}\right)\phi_{NL}(t,{{\bf x}})={\bf \Psi_L}\phi_{NL}\label{ansatzD} \eeqa where $\phi_{NL}(t,{{\bf x}})$ is a Lorentz scalar. We find
%If we solve the non-linear Dirac equation
%\beqa i\hbar \partial_t {\bf \Psi}(t,{\bf x})-{\bf \alpha}c{\hbar\over i}{\bf \bigtriangledown}{\bf \Psi}(t,{\bf x})-(mc^2+V_L+V_{NL})\beta {\bf \Psi}(t,{\bf x})=0\label{Dirac}\eeqa  in the same way that we solved the Schroedinger equation, imposing again that ${\bf \Psi_L}$  fulfills the linear Dirac equation, 

\beqa (\phi_{NL}(t,{{\bf x}}))(\beta i\hbar \partial_t {\bf \Psi^L}(t,{\bf x})-\beta{\bf \alpha}c{\hbar\over i}{\bf \bigtriangledown}{\bf \Psi^L}(t,{\bf x})-(mc^2+V_L){\bf \Psi^L}(t,{\bf x}))\nonumber\\
+(\beta {\bf \Psi^L}(t,{\bf x}) {i\hbar \partial \phi_{NL}(t,{{\bf x}})\over \partial t} -\beta{\bf \alpha}c {\bf \Psi^L}(t,{\bf x}){\hbar \over i}{\bf \bigtriangledown}\phi_{NL}(t,{{\bf x}}) \nonumber\\
 +{\bf \Psi^L}(t,{\bf x})V_{NL}(|{\bf \Psi^L}\phi_{NL}|) \phi_{NL}(t,{{\bf x}}))=0,\eeqa    

 that we solve in the same way as in the section 2 by requiring two constraints to be fulfilled:
 
 \beqa\beta i\hbar \partial_t {\bf \Psi^L}(t,{\bf x})-\beta{\bf \alpha}c{\hbar\over i}{\bf \bigtriangledown}{\bf \Psi^L}(t,{\bf x})-(mc^2+V_L){\bf \Psi^L}(t,{\bf x})=0,\eeqa which is  is the usual, linear, Dirac equation, and the non-linear constraint
 
  \beqa\beta {\bf \Psi^L}(t,{\bf x}) {i\hbar \partial \phi_{NL}(t,{{\bf x}})\over \partial t} -\beta{\bf \alpha}c {\bf \Psi^L}(t,{\bf x}){\hbar \over i}{\bf \bigtriangledown}\phi_{NL}(t,{{\bf x}}) \nonumber\\+{\bf \Psi^L}(t,{\bf x})V_{NL}(|\Psi(t,{{\bf x}})|_4) \phi_{NL}(t,{{\bf x}}))=0.\label{unknwonw}\eeqa

 In agreement with (\ref{Coulomb}), $V_{NL}$ supposedly scales like a Newtonian self interaction:
 %$ \phi_G^{self}(\lambda \Psi(t,{{\bf x}})) $=$ |\lambda|^2| \phi_G^{self}(\lambda \Psi(t,{{\bf x}})) $ and 

\beqa V_{NL}({\bf \Psi^L}\phi_{NL}) =A_L^2V_{NL}(({\bf \Psi^L}(t,{\bf x})/A_L)\cdot \phi_{NL}(t,{{\bf x}})) \phi_{NL}(t,{{\bf x}})\label{scalingV}\eeqa 

%where $|\Psi(t,{{\bf x}})|_4$ is the local density of stuff: 
where

%$|\Psi(t,{{\bf x}})|^2_4=|A_L|^2|\phi_{NL}(t,{{\bf x}})|^2=<{\bf \Psi^L}|_4{\bf \Psi^L}>(t,{{\bf x}})|\phi_{NL}(t,{{\bf x}})|^2$  with 
\beqa A_L=\sqrt{| \Psi^L_0(t,{{\bf x}})|^2+| \Psi^L_1(t,{{\bf x}})|^2+| \Psi^L_2(t,{{\bf x}})|^2+| \Psi^L_3(t,{{\bf x}})|^2}.\eeqa

At this level, a serious problem appears: equation (\ref{unknwonw}) consists of four equations. In the non-relativistic limit \cite{Messiah} and in absence of external magnetic field, spin decouples and can be factorized, but in general the four equations implicitly contained in equation (\ref{unknwonw}) are NOT equivalent to each other. 

A posteriori, we feel free to formally tackle the problem by proceeding as follows. We firstly decompose ${\bf \Psi}$ into its projection along ${\bf \Psi_L}$ and its projection along the 4-spinors orthogonal to ${\bf \Psi_L}$:
\beqa {\bf \Psi}={\bf \Psi_L}({\bf \Psi_L}^\dagger  {\bf \Psi})+(1-{\bf \Psi_L}\cdot {\bf \Psi_L}^\dagger ) {\bf \Psi}\label{ansatzDgen}\eeqa

Let us introduce $\phi_{NL}(t,{{\bf x}})$ through ${\bf \Psi_L}({\bf \Psi_L}^\dagger  {\bf \Psi})$ = ${\bf \Psi_L}\phi_{NL}(t,{{\bf x}})$, as well as the auxillary spinor $ {\bf \delta \Psi}$=$(1-{\bf \Psi_L}\cdot {\bf \Psi_L}^\dagger ) {\bf \Psi}$, which allows us to rewrite (\ref{ansatzDgen}) as follows:
\beqa {\bf \Psi}={\bf \Psi_L}\phi_{NL}+ {\bf \delta \Psi} \label{ansatzDdoublegen} \eeqa

 In a second time, let us simply neglect ${\bf \delta \Psi}$, in a grossly coarse-grained approach, in order to gain more insight about the physics of the problem, but keeping in mind however that there is a serious difficulty hidden under the rug at this level.

Retrospectively, our approximation justifies the ansatz (\ref{ansatzD}).
4-multiplying (\ref{unknwonw}) by $({\bf \Psi^L}(t,{\bf x}))^\dagger \beta$ and dividing by  $({\bf \Psi^L}(t,{\bf x}))^\dagger \cdot_4 {\bf \Psi^L}(t,{\bf x}))=A_L^2$ we get

  \beqa{i\hbar \partial  \phi_{NL}(t,{{\bf x}})\over \partial t}=\label{guidance''}         {\bf v^L_{Dirac}}(t,{{\bf x}}){\hbar \over i}{\bf \bigtriangledown} \phi_{NL}(t,{{\bf x}})\\ \nonumber +<\beta>_4 \cdot  A_L^2\cdot V_{NL}(\phi_{NL}) \phi_{NL}(t,{{\bf x}}). \eeqa
 where $<\beta>_4$=$({\bf \Psi^L}(t,{\bf x}))^\dagger \beta {\bf \Psi^L}(t,{\bf x}))$/$({\bf \Psi^L}(t,{\bf x}))^\dagger \cdot_4 {\bf \Psi^L}(t,{\bf x}))$
while ${\bf v^L_{Dirac}}$ obeys 
\beqa {\bf v^L_{Dirac}}\equiv{({\bf \Psi^L})^\dagger \alpha c {\bf \Psi^L}\over ({\bf \Psi^L})^\dagger {\bf \Psi^L}}\label{exact}\eeqa

 As is well-known, the conservation equation associated to the linear Dirac equation is ${\partial A^2_L\over \partial t}=-div(A_L^2 \cdot   {\bf v_{Dirac}}(t,{{\bf x_0}}))$. The dB-B guidance equation is thus nothing else than (\ref{exact}) \cite{Takaba,Holland}. Now, if we suppress in (\ref{S2}) the kinetic energy and the factor proportional to $A_L$, we find an equation which is formally equivalent to (\ref{guidance''}). Therefore, by repeating computations similar to those made in section 2, one can easily show that the dB-B guidance equation linked to Dirac's equation (\ref{exact}) is exactly satisfied.
Furthermore, resorting to the conservation equation ${\partial A^2_L\over \partial t}=-div(A_L^2 \cdot   {\bf v_{Dirac}}(t,{{\bf x_0}}))$, it is straightforward to establish the validity of property 2 as in the non-relativistic case studied before. For instance ${{d A_L\over dt}\over A_L}={-1\over 2}{1\over <\phi_{NL}|\phi_{NL}>}{d <\phi_{NL}|\phi_{NL}>\over dt}$

 Consequently, we are entitled to look for peaked wave functions of the form

\beqa {\bf \Psi}=\left(\begin{array}{c}
 \Psi_0(t,{{\bf x}}) \\
 \Psi_1(t,{{\bf x}}) \\
 \Psi_2(t,{{\bf x}}) \\
 \Psi_3(t,{{\bf x}}) \\
\end{array}\right) ={1\over A_L}\cdot \left(\begin{array}{c}
 \Psi^L_0(t,{{\bf x}}) \\
 \Psi^L_1(t,{{\bf x}}) \\
 \Psi^L_2(t,{{\bf x}}) \\
 \Psi^L_3(t,{{\bf x}}) \\
\end{array}\right)\cdot \phi'_{NL}(t,{{\bf x}}),\eeqa
for which we know that

(Property 1) the barycentre of  $ \phi'_{NL}$ moves along the hydrodynamical flow lines of Dirac's linear equation, at velocity ${\bf v_{Dirac}}$.

(Property 2) the L$_2$ norm of $ \phi'_{NL}(t,{{\bf x}})$ is constant throughout time.

 %In particular, if $A_L$ and ${\bf v^0_{Dirac}}$ are smooth enough, we find at the lowest order of approximation  the solution
%$ \phi'_{NL}(t,{{\bf x}})\approx e^{-iE_0t/\hbar}\cdot \phi^{0}_{NL}({{\bf x}}-{{\bf x^0}}(t=0)-\int^t_{0} dt {\bf v^0_{Dirac}})$

%where $E_0=<\beta>_4 mc^2$ and \beqa \Psi^0=e^{-i<\beta>_4 mc^2t/\hbar}({1\over A_L}){\bf \Psi^L}(t,{{\bf x}})\cdot \phi^0_{NL}({\bf x}-{\bf x_0(t)}),\label{approsinglenew}\eeqa

As in the non-relativistic case, the amplitude $A_L$ is an auxiliary function that disappears at the end of the computation of ${\bf \Psi}$. 
It is worth noting that we did not demonstrate the existence of a ground state of the type $\phi^{0}_{NL}$ in the relativistic case\footnote{\label{footnotezitter}Setting $V_L=0$ in (\ref{NLD})  an mutiplying it by $(\beta i\hbar \partial_t -\beta{\bf \alpha}c{\hbar\over i}{\bf  \bigtriangledown}-mc^2(1+\phi_G^{self}/c^2))$ where we denoted $\phi_G^{self}=-Gm\int {d}^3 x'(\frac{|\Psi(t,{{\bf x'}})|^2}{|{{\bf x} -{\bf x'}}|})$, we get the non-linear Klein-Gordon equation
\beqa \hbar^2c^2({1\over c^2}({\partial^2\over \partial t^2}-\Delta)) {\bf \Psi}(t,{\bf x})   +  m^2c^4(1+\phi_G^{self}/c^2)^2{\bf \Psi}(t,{\bf x})  =0        \eeqa 
Then the static version of the non-linear Klein-Gordon equation reads \beqa -\hbar^2c^2\Delta {\bf \Psi}({\bf x})   +  m^2c^4(1+\phi_G^{self}/c^2)^2{\bf \Psi}({\bf x})  =\hbar^2 E^2    {\bf \Psi}({\bf x})    \label{KGNL}    \eeqa In the weak potential limit ($\phi^{self}_G<<c^2$) the static solution  of (\ref{KGNL}) also satisfies Choquard equation (\ref{Choquard}), up to rescaling. In the strong field limit, we have no guarantee about the existence of such a solution, but if it exists then the non-linear Klein-Gordon equation and its static counterpart, the non-linear Choquard equation constitute an interesting toy-model for self-gravity, even in the relativistic regime. We suspect however the appearance of difficulties in the same limit, related to zitter bewegung \cite{Messiah,Hestenes}, because in general the spin does not decouple from $\phi_{NL}$ in (\ref{unknwonw}).}. We also suspect that the self-gravitational potential $V_{NL}$ proposed in (\ref{Newton}) is not the right candidate in a relativistic context because it is obviously not Lorentz covariant. In any case, it is beyond the scope of our paper to characterize in detail what ``really'' happens at the level of the soliton. We expect for instance \cite{new} that in the case of an electron, the size of the soliton is of the order of 10$^{-57}$ meter, well beyond the Planck scale. Our main aim in the present section was merely to suggest that hopefully properties 1 and 2 exhibit some structural invariance when we pass from the non-relativistic to the relativistic regime.

\end{document}